\documentclass[aps,preprint,superscriptaddress,showpacs]{revtex4}
\usepackage{graphicx}
\usepackage{dcolumn}
\usepackage{amsmath}

\begin{document}
\bibliographystyle{apsrev}

\title{Acceleration effect caused by the Onsager reaction term
in a frustrated coupled oscillator system
} 

\author{Toru Aonishi}

\affiliation{Brain Science Institute, RIKEN, 
2-1 Hirosawa, Wako-shi, Saitama, 351-0198, Japan 
}
\author{Masato Okada}

\affiliation{Brain Science Institute, RIKEN, 
2-1 Hirosawa, Wako-shi, Saitama, 351-0198, Japan 
}

\affiliation{ERATO Kawato Dynamic Brain Project, 2-2 Hikaridai, Seika-cho, 
Soraku-gun, Kyoto 619-0288, Japan}

\date{\today}

\begin{abstract}
The role of the {\it Onsager reaction term} (ORT) 
is not yet well understood in frustrated coupled oscillator systems,
since the Thouless-Anderson-Palmer (TAP) and 
replica methods cannot be directly applied to these non-equilibrium systems.
In this paper, 
we consider two oscillator associative memory models,
one with symmetric and one with asymmetric dilution of coupling.
These two systems are ideal for evaluating the effect of the ORT,
because, with the exception of the ORT, 
they have the same order parameter equations.
We found that the two systems have identical macroscopic properties,
except for the acceleration effect caused by the ORT.
This acceleration effect does not exist in any equilibrium system.
\end{abstract}


\maketitle

Coupled oscillators are of intrinsic interest in many branches of
physics, chemistry and biology.
Simple coupled-oscillator models involving uniform and global coupling
have been investigated in some detail, and it has been found that 
they can be used to model many types of chemical reactions in solution \cite{kuramoto0}.
However, in the modeling of more complicated phenomena, 
including those studied in the 
fields of neuronal systems, it is more necessary
to consider coupled oscillators with frustrated couplings.

The {\it Onsager reaction term} (ORT), which describes the effective
self-interaction, is of great importance in obtaining a physical  
understanding of frustrated random systems, because the 
presence of such an effective self-interaction is one of 
the characteristics that distinguish frustrated and non-frustrated
systems. In the case of equilibrium systems,
we can rigorously evaluate the effect of the ORT in 
the Thouless-Anderson-Palmer (TAP) framework \cite{Mezard},
and/or using the replica method \cite{fukai2}.
However, we cannot directly apply these systematic methods 
to non-equilibrium coupled-oscillator systems.
We can define a formal Hamiltonian function on such systems.
However, Perez et al. proved that the ground states 
of this Hamiltonian are not stationary states of the dynamics \cite{Perez}.
Therefore, it is impossible to construct a theory based on 
a free energy for this class of systems.
For this reason, in order to evaluate the macroscopic quantities 
in such systems that include an ORT, the 
{\it self-consistent signal to noise analysis} (SCSNA)
which can be applied to systems without the Hamiltonian function   
has been used \cite{fukai3}.
The mathematical treatment of this method is similar to that of the
cavity method \cite{Mezard}.
Some results obtained from the SCSNA consist with results from 
the replica method, but this method includes a few heuristic steps.
While the SCSNA has indicated some interesting results, it is not 
sufficient to give a complete understanding of frustrated systems,
and for this reason many theoretically fundamental questions remain 
in the study of such systems. In fact, even the existence of the type of
self-interaction that can be described by the ORT 
is the subject of some debate \cite{aonishi2,yoshioka}. 

In this paper, we discuss an effect of the ORT that exists only 
in frustrated globally coupled oscillator systems,
and in particular cannot be found in equilibrium systems.
In order to make this effect clear,
it would be ideal for us to compare two frustrated systems
that, with the exception of the different quantity of the ORT, 
have the same order parameter equations.
In addition, it is desirable for  these systems to have
a clear correspondence with an equilibrium system,
because the effects of the ORT are well understood in equilibrium systems.

In consideration of the above-mentioned points, a system of the form:
\begin{eqnarray}
\frac{d \phi_i}{dt} = \omega_i 
 + \sum_{j \neq i}^{N} J_{ij}\sin(\phi_j - \phi_i + \beta_{ij} + \beta_0), 
 \label{Eq.model}
\end{eqnarray}
is ideal. In fact, such systems are well known as models 
of coupled oscillator systems \cite{kuramoto0,Park}.
Here, $\phi_i$ is the phase of the $i$ th oscillator 
(with a total of $N$) and $\omega_i$ represents its quenched 
natural frequency.
The natural frequencies are randomly distributed 
with a density represented by $g(\omega)$.
We restrict $g(\omega)$ to a unimodal symmetric distribution for
satisfying the condition that there exists one large cluster of
synchronous oscillators.
Also in Eq. (\ref{Eq.model}), $J_{ij}$ and $\beta_{ij}$ denote the
amplitude of coupling from unit $j$ to unit $i$ 
and its delay, respectively.
In the present study, we have selected the following two
generalized Hebb learning rules with random dilutions \cite{aoyagi2}
to determine $J_{ij}$ and $\beta_{ij}$:
\begin{eqnarray}
& &J_{ij} \exp(i \beta_{ij}) 
  = \frac{c_{ij}}{c N} \sum_{\mu=1}^{p}
                 \xi_{i}^{\mu} {\overline\xi}_{j}^{\mu}, \quad
		 \xi_i^{\mu} = \exp( i \theta_i^{\mu}), \\ 
& &c_{ij} = \left\{ \begin{array}{ll}
1 \quad & {\rm with\  probability}\ c\\
0 \quad & {\rm with\  probability}\ 1-c
\end{array} \right. , 
\end{eqnarray}
where $\overline{\cdot}$ means the complex conjugate. 
$\{ \theta_i^\mu \}_{i=1, \cdots, N, \mu=1, \cdots, p}$
are the phase patterns to be stored in the present model and are
assigned to random numbers with a uniform probability on the interval $[0,
2\pi)$. $\mu$ is an index of stored patterns
and $p$ is the total number of stored patterns. 
We define a parameter $\alpha$ (the loading rate) by 
$\alpha=p/N$. When $\alpha \sim O(1)$, 
the system has frustration.
The quantity $c_{ij}$ is the dilution coefficient.
Let $c_{ij}=1$ if there is a non-zero coupling from unit $j$ to 
unit $i$ and $c_{ij}=0$ otherwise.
The number of fan-in (fan-out) is restricted to $O(N)$, 
i.e., $c \sim O(1)$.
Here, we consider both the cases of symmetric dilution 
(i.e., $c_{ij}=c_{ji}$) and
asymmetric dilution (i.e., $c_{ij}$ and $c_{ji}$ are independent
random variables)\cite{okada}.  The quantity $\beta_0$ in 
Eq. (\ref{Eq.model}) represents a uniform bias.
Due to the effect of this bias, the mutual interaction between a pair of 
oscillators is asymmetric, even if $J_{ij}= J_{ji}$ and $\beta_{ij} =
-\beta_{ji}$.
Such an unbalanced mutual interaction is the essence of 
the acceleration (deceleration) effect \cite{meunier,Hansel1,kurata}.
In the case of $g(\omega) =\delta(\omega - \omega_0)$, $\beta_0=0$
and $c_{ij}=c_{ji}$,
this system can be mapped to a XY-spin system \cite{aoyagi2,cook,okuda}. 
In this way, we can make a bridge between 
the frustrated coupled oscillator system
and the equilibrium system.

Let us consider steady states of the system in the limit 
$t \rightarrow \infty$.
Our theory is based on the condition that there exists one large cluster of oscillators 
synchronously locked at frequency $\Omega$ and the number
of this cluster scales as $\sim$ $O(N)$. 
Under such a condition, Daido demonstrated through a scaling plot obtained from 
numerical simulations that variation of order parameter scales as $O(1/\sqrt{N})$ 
in ferromagnetic systems with one large synchronous cluster \cite{daido2}. 
According to this result, we assume that the self-averaging property holds 
in our system and order parameters are constant in the limit $N
\rightarrow \infty$. These assumptions in our theory were also introduced by 
Sakaguchi and Kuramoto (SK) \cite{kuramoto}.

Redefining  $\phi_i$ according to $\phi_i \rightarrow  \phi_i + \Omega
t$ and substituting this into Eq. (\ref{Eq.model}), we obtain 
\begin{eqnarray}
- \frac{d \phi_i}{dt} + \omega_i - \Omega = \sin(\phi_i) h^R_i - \cos(\phi_i) h^I_i \label{eq:hei1},
\end{eqnarray}
where $h_i$ represents the so-called ``local field'',
which is described as
\begin{eqnarray}
& &h_i = h_i^R +i h_i^I  = \nonumber \\
& &e^{i \beta_0}
\left( \sum_{\mu}^{p} \xi_i^\mu m^\mu + \frac{1}{N}\sum_{\mu}^{p}
\sum_{j\neq i}^{N} 
\frac{c_{ij}-c}{c} \xi_{i}^{\mu} {\overline\xi}_{j}^{\mu} s_j -
\alpha s_i\right).
\label{eq:lf1}
\end{eqnarray}
For convenience, we write $s_i = \exp(i \phi_i)$.
The order parameter $m^\mu$, which is the overlap between 
the system state $\{ s_i \}_{i=1, \cdots, N}$ 
and embedded pattern $\{ \xi_i^\mu \}_{i=1, \cdots, N}$, is defined as 
\begin{eqnarray}
m^\mu = \frac{1}{N} \sum_{j=1}^{N} {\overline\xi}_j^\mu s_j.
\label{eq:ov1}
\end{eqnarray}
The effect of the second term of Eq. (\ref{eq:lf1}),
i.e., $\frac{1}{N}\sum_{\mu}^{p} \sum_{j\neq i}^{N} 
\frac{c_{ij}-c}{c} \xi_{i}^{\mu} {\overline\xi}_{j}^{\mu} s_j$,
is equivalent to 
that of an effect of additive coupling noise \cite{aoyagi2,Sompolinsky2,okada}.
In the limit $c \rightarrow 0$, with $\alpha /c$ kept finite, 
our system is reduced to a glass oscillator.
Therefore, our theory proposed here can cover two types of frustrated
systems, the oscillator associative memory and the glass oscillator. 
 
In general, the fields $h^R_i$ and $h^I_i$ involve the ORT
corresponding to the effective self-feedback \cite{fukai3}.
We must eliminate the reaction term from these fields.
Here, we assume that the local field splits into a ``pure'' effective local
field, $\tilde{h}_i = \tilde{h}^R_i + i \tilde{h}^I_i$, and the ORT, $\Gamma s_i$:
\begin{eqnarray}
h_i= \tilde{h}_i +  \Gamma s_i. \label{eq:as2} 
\end{eqnarray}
Here we have neglected the complex conjugate term of the ORT leading to
a higher-harmonic term of the response function \cite{aonishi}.
This can be done in the present model because we employ generalized Hebb learning rules. 
Hence, by substituting Eq. (\ref{eq:as2}) into
Eq. (\ref{eq:hei1}), we obtain the equation
\begin{eqnarray}
- \frac{d \phi_i}{dt} + 
\omega_i - \tilde{\Omega} = \sin(\phi_i) \tilde{h}^R_i -
\cos(\phi_i) \tilde{h}^I_i,  \label{eq:hei2} \\
\tilde{\Omega} = \Omega - \left|\Gamma \right|\sin(\psi), \ \
\psi = {\rm Arg} \left(\Gamma \right), \label{eq:gamma} 
\end{eqnarray}
which does not contain the reaction term.
The quantity $\tilde{\Omega}$ here represents the effective frequency of synchronous
oscillators.
We can regard $\tilde{\Omega}$ as the renormalized version of
$\Omega$, of which the ORT has been pulled out, and therefore 
$\tilde{\Omega}$ takes a different value 
from the observable $\Omega$ in general.
Thus, $\Omega - \tilde{\Omega}$ represents  
the contribution of the ORT 
to the acceleration (deceleration) effect.
$\tilde{\Omega}$ is one of the order parameters of our theory.
In the analysis that follows, $\tilde{h}_i$ and $\Gamma$ are obtained 
in a self-consistent manner. 
Under the above assumption expressed by Eq. (\ref{eq:as2}), by applying SK 
theory \cite{kuramoto} to Eq. (\ref{eq:hei2}),
we obtain the average of $s_i$ over $\omega_i$:
\begin{eqnarray}
\left<s_i \right>_{\omega_i} &=& \tilde{h_i} \int_{-\pi/2}^{\pi/2} d \phi g\left(\tilde{\Omega} +
|\tilde{h}|\sin\phi \right)\cos\phi\exp(i \phi)\nonumber \\ 
&+& i  \tilde{h}_i \int_{0}^{\pi/2} d \phi 
\frac{\cos\phi (1-\cos\phi)}{\sin^3\phi} \nonumber \\ 
& &\hspace{0.3cm}\times \left\{ 
g\left(\tilde{\Omega}+\frac{|\tilde{h}_i|}{\sin\phi}\right)-g\left(\tilde{\Omega}-\frac{|\tilde{h}_i|}{\sin\phi}\right)
\right\}. \label{Eq.X}
\end{eqnarray}

Equation (\ref{eq:hei1}) implies that $s_i$ is a function of $h_i$,
$\omega_i-\Omega$ and $t$, and to make this explicit we write 
\begin{equation}
 s_i = X(h_i, \omega_i-\Omega, t).
\end{equation}
Note that $s_i$ is not a function of 
renormalized $\tilde{h}_i$ and $\tilde{\Omega}$
but, rather, the bare $h_i$ and $\Omega$ in appearing Eq. (\ref{eq:hei1}).
We can properly evaluate the ORT with this careful treatment.
Here, we assume that microscopic memory effect can be neglected
in the $t \rightarrow \infty$ limit. 
In this analysis, we focus on memory retrieval states in which the
configuration has appreciable overlap with the condensed pattern 
$\mbox{\boldmath $\xi$}^1$ ($m^1 \sim O(1)$)
and has tiny overlap with the uncondensed patterns 
$\mbox{\boldmath $\xi$}^\mu$ for $\mu>1$ ($m^\mu \sim O(1/\sqrt{N})$).
Under this assumption, we estimate the contribution of 
the uncondensed patterns using the SCSNA\cite{fukai3},
and determine $\tilde{h}_i$ in a self-consistent manner.
Finally, the equations relating the order parameters $|m^1|$, $U$ and $\tilde{\Omega}$
are obtained using the self-consistent local field:
\begin{eqnarray}
|m^1| e^{-i\beta_0}&=& \left<\left< \tilde{X}(x_1,x_2;\tilde{\Omega}) \right>\right>_{x_1, x_2}, \label{Eq.o1} \\
U e^{-i\beta_0}&=& \left<\left< F_1(x_1,x_2;\tilde{\Omega}) \right>\right>_{x_1, x_2}, \label{Eq.o2}
\end{eqnarray}
where $\left<\left< \cdots \right>\right>_{x_1, x_2}$ is 
the Gaussian average over $x_1$ and $x_2$,
$\left<\left< \cdots \right>\right>_{x_1, x_2} = \int
\int Dx_1 Dx_2\cdots$. The quantity $U$ corresponds to the susceptibility,
which is the measure of the sensitivity to external fields.
Since the present system possesses rotational symmetry
with respect to the phase $\phi_i$,
we can safely set the condensed pattern as $\xi^1_i=1\ (i=1,\cdots, N)$.
Now, $\tilde{h}$, $\tilde{X}$, $F_1$ and $Dx_1 Dx_2$ 
can be expressed as follows:
\begin{eqnarray}
& &Dx_1 Dx_2=\frac{dx_1 dx_2}{2 \pi \rho^2} 
\exp\left(-\frac{x_1^2+x_2^2}{2 \rho^2} \right),\\
& &\rho^2 = \frac{\alpha}{2}\left(\frac{1}{|1-U|^2} +
\frac{1-c}{c}\right), \\
& & \tilde{h} = |m^1| + x_1+i x_2, \\
& &\tilde{X}(x_1,x_2;\tilde{\Omega}) = \left<s \right>_\omega, \label{eq:aa} \\
& &F_1(x_1,x_2;\tilde{\Omega})= \frac{\partial \left<s \right>_\omega}{\partial \tilde{h}}, \label{eq:bb}
\end{eqnarray}
where $\left<s \right>_\omega$ in Eqs. (\ref{eq:aa}) and (\ref{eq:bb})
is written as Eq. (\ref{Eq.X}).

In the case of the symmetric diluted system, $\Gamma$ can be expressed as 
\begin{eqnarray}
\Gamma e^{-i\beta_0} = \frac{\alpha U}{1-U} + \frac{\alpha (1-c)}{c} U. \label{eq:ga1}
\end{eqnarray}
In the case of the asymmetric diluted system, on the other hand, we have
\begin{eqnarray}
\Gamma e^{-i\beta_0} = \frac{\alpha U}{1-U}. \label{eq:ga2}
\end{eqnarray}

$\tilde{h}$ and $\tilde{\Omega}$ are the renormalized versions of
$h$ and $\Omega$ respectively, of which the ORT has been pulled out, 
and thus $\tilde{h}$ and $\tilde{\Omega}$ are independent of 
the ORT. Therefore, the two models 
we consider have identical order parameter 
equations, (\ref{Eq.o1}) and (\ref{Eq.o2}), written  in the term of 
the renormalized quantities $\tilde{h}$ and $\tilde{\Omega}$.
According to Eq. (\ref{eq:gamma}), the difference of the ORT in Eqs. (\ref{eq:ga1}) and (\ref{eq:ga2}) 
only leads a different value of the observable $\Omega$ only 
when $\beta_0 \neq 0$.
In this way we are able to clearly separate the effect of the ORT,
and therefore, by observing the macroscopic parameter $\Omega$ of these
two systems, we can analyze the effect of the ORT
qualitatively and quantitatively. 

The distribution of resultant frequencies $\overline{\omega}$ in the
memory retrieval state, which is denoted 
$p(\overline{\omega})$, becomes 
\begin{eqnarray}
& &p(\overline{\omega}) = r
\delta(\overline{\omega}-\Omega) \nonumber \\
& &+
\int Dx_1 Dx_2\frac{g\left( \tilde{\Omega}+ (\overline{\omega}- \Omega)\sqrt{1 + 
\frac{|\tilde{h}|^2}{(\overline{\omega}-\Omega)^2}}\right)}{
\sqrt{1 +
\frac{|\tilde{h}|^2}{(\overline{\omega}-\Omega)^2}}}, \label{eq:dis} \\
& &r =\int Dx_1 Dx_2 |\tilde{h}| \int_{-\pi/2}^{\pi/2} d\phi
g\left(\tilde{\Omega} + |\tilde{h}| \sin\phi \right) \cos\phi .
\end{eqnarray}
The $\delta$-function in Eq. (\ref{eq:dis}) indicates the cluster of 
oscillators synchronously locked at frequency $\Omega$.
The value $r$ is the ratio between the number of synchronous oscillators
and the total number of oscillators $N$.
The second term in Eq. (\ref{eq:dis}) represent  a distribution of asynchronous oscillators.

If $\beta_0=0$ and $g(\omega)$ is symmetric, our theory reduces to
the previously proposed theory \cite{aonishi2}.
In the case $g(\omega)=\delta(\omega)$, $\beta_0 = 0$ and
$c_{ij}=c_{ji}$, where the present model reduces to 
XY-spin systems (frustrated equilibrium systems), 
our theory coincides with the replica theory \cite{aoyagi2,cook} 
and the SCSNA\cite{okuda}.
In addition, in the limit of $\alpha \rightarrow 0$, 
where the present model reduces to uniform non-equilibrium systems,
our theory reproduces the SK theory \cite{kuramoto}.

\begin{figure}[ht]
\includegraphics[height=7cm]{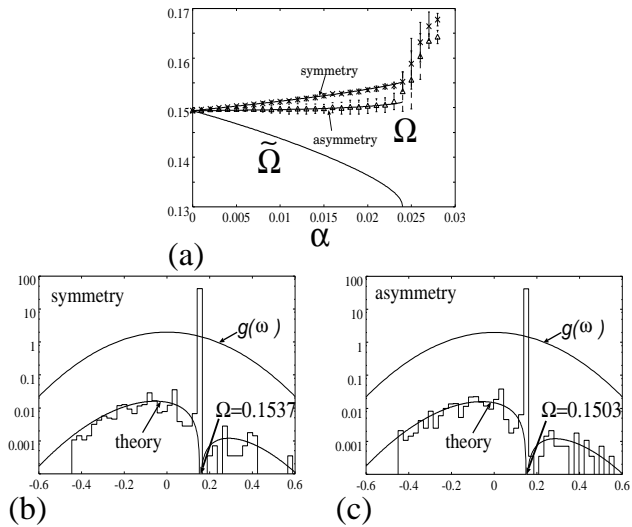}
\caption{ The difference between symmetric and asymmetric dilution
systems. Here $N=10,000$, $\sigma=0.2$, $\beta_0=\pi/20$, and $c=0.5$.
(a) $\Omega$ as a function of $\alpha$. 
(b) The distribution of resultant frequencies for the symmetric dilution
system ($\alpha=0.02$).
(c) The distribution of resultant frequencies for the asymmetric dilution
system ($\alpha=0.02$).
}
\label{tho2}
\end{figure}

\begin{figure}[ht]
\includegraphics[height=3.5cm]{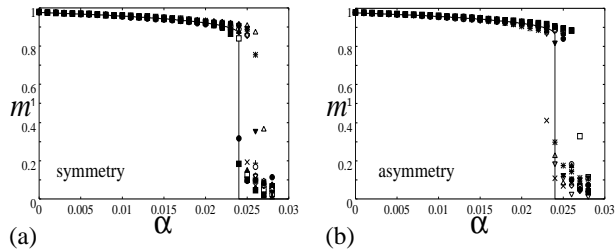}
\caption{$|m^1|$ as a function of $\alpha$. The solid curves were
obtained theoretically, and the data points were obtained from numerical
simulations. Here $N=10,000$, $\sigma=0.2$, $\beta_0=\pi/20$, and $c=0.5$.
(a) Symmetric dilution system. (b) Asymmetric dilution system.
}
\label{overlap}
\end{figure}

In the numerical simulations we now discuss, we chose 
the distribution of natural frequencies as 
$g(\omega)=(2\pi\sigma^2)^{-1/2} \exp(-\omega^2/2\sigma^2)$.
Also, we set $\sigma=0.2$, $\beta_0=\pi/20$, and $c=0.5$.
Figure \ref{tho2}(a) displays  $\Omega$ as a function of  $\alpha$ 
in the memory retrieval states, where the solid
curves were obtained theoretically, and the data points with error bars 
represent results obtained by 
numerical simulation.
As seen from Figure \ref{tho2}(a), 
the oscillator's rotation for the symmetric diluted 
system is faster than that for the asymmetric diluted system.
As previously discussed, 
$\tilde{\Omega}$ in Figure \ref{tho2}(a),
which represents the effective frequency of synchronous oscillators, 
does not depend on the type of dilution,
while the observed $\Omega$ depends strongly on it.
This dependence is due to the existence of the ORT $\Gamma$.
In the case that the local field $h$ does not contain the ORT \cite{yoshioka}, 
the plots obtained from the  numerical simulations
for both models should fit the curve of $\tilde{\Omega}$.
Therefore, the dependence of the observed $\Omega$ on the type of dilution
is strong evidence for the existence of the ORT
in the present system.
In this figure, we have shifted the numerical values of $\Omega$ at $\alpha=0$
(in the computer simulation) 
to their corresponding theoretical values at $\alpha=0$,
in order to cancel fluctuations of  
the mean value of $g(\omega)$ caused by the finite size effect. 

Figures \ref{tho2}(b) and (c) display
the distributions of the resultant frequencies 
for the symmetric dilution and asymmetric dilution systems,
respectively. 
As these figures reveal, the theory (solid curve) is in good agreement 
with the simulation results (histogram).
According to the results given in Figures \ref{tho2}(b) and (c), 
the distributions of the resultant frequencies for the symmetric diluted
system is identical to that for the asymmetric diluted system,
except for a slight difference between those positions caused by 
the ORT. 
From this result we can conclude that the mean field $\tilde{h}$ 
of the symmetric diluted system 
is identical to that of the asymmetric diluted system,
since $\tilde{h}$ reflects 
the distribution of resultant frequencies, 
as represented by Eq. (\ref{eq:dis}).
Figures \ref{overlap}(a) and (b) display $|m^1|$ as a function of  $\alpha$
for symmetric and asymmetric diluted systems,
respectively. Here, the solid curves were obtained theoretically, 
and the data plots represent the results obtained from the numerical simulations.
According to the results given in Figures \ref{overlap}(a) and (b),
the critical memory capacity of 
the asymmetric diluted systems obtained from numerical simulation is
slightly smaller than that of the asymmetric diluted systems,
because asymmetric dilutions break the detailed balance, and then, 
asymmetric dilutions weaken a memory state. At this stage, there is no theory to 
rigorously treat a system with asymmetric interaction.
Almost theoretical studies of asymmetric systems 
are based on the naive assumption that there exist steady states 
like symmetric systems \cite{okada}. 

In conclusion, we have found that the symmetric and asymmetric diluted systems have 
the same macroscopic properties, with the exception of the 
the acceleration (deceleration) effect caused by the ORT. 
The quantity of the ORT depends on the type of dilution, and this dependence leads to a difference 
in the rotation speed of the oscillators for the two cases, as shown in Fig. \ref{tho2}(a).
The acceleration (deceleration) effect caused by the ORT 
is a phenomenon peculiar to non-equilibrium systems,
since this effect only exists when $\beta_0 \neq 0$.
There has been fundamental disagreement regarding the existence 
of the ORT in a typical system
corresponding to our model with $\beta_0=0$ and  a symmetric $g(\omega)$
\cite{aonishi2,yoshioka}. 
In this work we have reached the following conclusion in this regard.
Even if $\beta_0=0$ and $g(\omega)$ is symmetric, 
the ORT exists in the bare local field given by Eq. (\ref{eq:lf1}).
In this case, the effect of the ORT is invisible,
since it cancels out of Eq. (\ref{eq:hei2}).

\bibliography{ref.bib}

\end{document}